# Theoretical Limits and Scaling Laws for Electrokinetic Molecular Concentration via Ion Concentration Polarization

Wei Ouyang,[1] Zirui Li,[2,*] Xinghui Ye,[2] Jongyoon Han[1,2,3,*]

[1]Department of Electrical Engineering and Computer Science, Massachusetts Institute of Technology, Cambridge, Massachusetts, 02139, USA.

[2]Institute of Laser and Optoelectronic Intelligent Manufacturing, College of Mechanical and Electrical Engineering, Wenzhou University, Wenzhou, 325035, P.R. China.

[3]Department of Biological Engineering, Massachusetts Institute of Technology, Cambridge, Massachusetts, 02139, USA.

We develop the first theoretical model for the analytical description of ion concentration polarization (ICP)-based electrokinetic molecular concentration, which had not been possible due to the extraordinary complexity of the system. We define the two separate limits for the enrichment factor achievable in a given system and derive the scaling laws for critical parameters, which are validated by numerical simulations and experiments. This work provides clear theoretical explanations on the diverse experimental behaviors previously observed yet unexplainable, while setting solid foundation for the engineering of ICP-based concentrators and other fluid-coupled electrokinetic systems.

Over the past decade, electrokinetics in micro-nanofluidic systems has been an unprecedentedly active research area with the aid of the recent advancements in nanotechnology [1]. One of the most exciting yet least understood systems in nanoscale electrokinetics is the electrokinetic molecular concentrator that is capable of million-fold concentration, which is enabled by the coupling of ion concentration polarization (ICP) at the micro-nanofluidic interfaces and net tangential electroosmotic flow (EOF) in a microchannel [2]. While extensive efforts have been devoted to developing ICP-based concentrators [2-6], understanding of the fundamental mechanisms is contrastingly poor. This is largely hampered by the extreme complexity of the system, which involves tightly coupled nonlinear fluid flow, ion and charged particle transport, and dynamic evolution of electric field in multi-scale space spanning from sub-nanometers to centimeters. Despite the significant research efforts and interests towards the understanding of componental problems, such as the ICP phenomenon [7-12], nonlinear electrokinetic flow [13-20], and transport of charged species in micro-nanofluidics [1,21,22], fully coupled analyses of the ICP-based electrokinetic concentration system are rarely reported. This resulted in the scientific difficulty in explaining the diverse experimental behaviors observed by different researchers [6], not to mention the inability to optimize and engineer the concentration performances.

Because of the strong coupling of diverse physical processes and the infeasibility in the direct measurement of some features (*e.g.* dynamics of buffer ions), accurate analysis of the electrokinetic concentration system has to rely on numerical simulations. In this regard, Jia *et al.* [23,24] numerically investigated the effects of the membrane charge density, the mobility of counter-ions through the CEM, and other parameters on the enrichment performance. In a more recent work, Li *et. al* [25] conducted a fully coupled simulation of the ICP-based concentration system, in which the mechanism of electrokinetic trapping is elaborated through analyzing the dynamics of ions and charged particles and the effects of system parameters were computed. Although the case-by-case simulation studies provide some key insights into the electrokinetic concentration system, analytical solutions, or at least scaling laws for key parameters, which enable the direction prediction of the enrichment factor (EF), are still highly desirable.

The aim of this letter is to derive the approximate analytical solutions of the EF of the ICP-based electrokinetic concentration system, establish the scaling laws for critical parameters, and compare them with simulation results and experimental data. From simulation results we identify two limits, namely the *electrokinetic* (*EK*) *limit* and the *electroneutrality* (*NT*) *limit*, for the maximum EF achievable in a given electrokinetic concentration system. Subsequently, we formulate the approximate analytical solutions of these two limits, based on which the scaling relations between these two limits and critical parameters are extracted. The resulted scaling laws are confirmed by numerical simulations and experimental data. These findings have practical implications in the engineering of molecular concentration systems.

Fig. 1(a) shows the schematic of the ICP-based concentrator. The selective transport of cations through a cation-exchange membrane (CEM) under an electric field $\mathbf{E_N}$ results in the formation of an ion depletion zone in the vicinity of the CEM in the upper microchannel. Meanwhile, the application of a tangential electric field $\mathbf{E_T}$ induces an electroosmotic flow along the microchannel. Negatively charged species are carried into the microchannel by the EOF and subsequently trapped at the front of the depletion zone by the locally amplified electric field. Fig. 1(b) shows the dual-channel ICP-based concentrator in the experiment, which was fabricated by polydimethylsiloxane (PDMS) soft lithography and plasma bonding [26]. The two microchannels are 2 cm long, 100 μm wide and 15 μm deep with a gap of 250 μm. The CEM (nafion, shown in purple) is 200 μm wide and ~1 μm thick. The default electrical configuration is $V_1$=20 V, $V_2$=10 V, $V_3$=10 V. The buffer is 10 mM KCl solution, and the analyte biomolecules are Alexa Fluor 488-labeled bovine serum albumin (BSA, carrying 14 negative charges at pH=7.4 [27]) and Alexa Fluor 488-labeled 21mer ssDNA (carrying 21 negative charges). As aforementioned, the biomolecules are injected into the channel following the EOF and trapped at the front of the depletion zone induced by the selective transport of ions through the CEM under $E_N$. The experimental details are described in Section 1 (S1) of the Supplemental Material (SM).

Fig. 1(c) shows the simulation model. The key component is a microchannel of 120 μm long and 4 μm wide connecting two reservoirs of 60 μm long and 60 μm wide. The microchannel walls are charged with a surface charge density of -5 mC/m$^2$. CEMs of 2 μm long are embedded in the middle of the top and bottom walls, which are assumed to permit the passage of cations only [12,18,28]. The default electrolyte is 1 mM potassium chloride (KCl) with 0.1 nM divalently negatively charged particles (representing the biomolecules to be concentrated) of a diffusion coefficient 1/4 of that of Cl$^-$. The left boundary and right boundary are biased to electric potentials of $V_L$=20$V_T$ and $V_R$=0, respectively. $V_T$=$kT/e$ is the thermal voltage, where $k$, $T$, $e$ are the Boltzmann constant, the absolute temperature, and the elementary charge, respectively. The CEMs are biased to an electric potential of $V_m$. In order to realize electrokinetic concentration, it is required that $V_L$>$V_R$ (generating net tangential EOF) and $V_{cm}$=($V_L$+$V_R$)/2-$V_m$>0 (generating ion depletion). For convenience, $V_{cm}$ is referred to as the *cross-membrane voltage*. The higher $V_{cm}$ is, the stronger the ion depletion effect is.

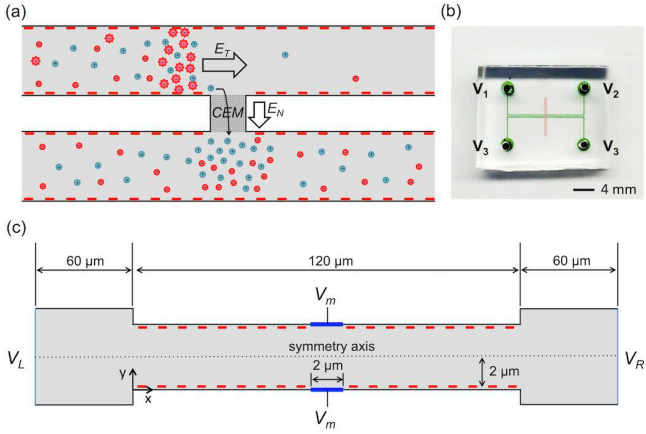

Fig. 1. (a) Schematic illustration of the dual-channel ICP-based concentrator. (b) Photo of the dual-channel concentrator. (c) The simulation model.

The governing equations for the incompressible fluid flow, ion and particle transport, and electric potential are the Navier-Stokes, Nernst-Planck, and Poisson equations [16], respectively:

$$\rho(\partial \mathbf{U}/\partial t + (\mathbf{U}\cdot\nabla)\mathbf{U}) = -\nabla P + \eta\nabla\cdot\nabla\mathbf{U} - \rho_e\nabla\Phi, \quad (1)$$

$$\nabla\cdot\mathbf{U} = 0, \quad (2)$$

$$\partial C_i/\partial t = -\nabla\cdot\mathbf{J}_i, \quad (3)$$

$$\mathbf{J}_i = -D_i\nabla C_i - D_i(Z_i e/kT)C_i\nabla\Phi + \mathbf{U}C_i, \quad (4)$$

$$-\nabla\cdot(\varepsilon\nabla\Phi) = \rho_e, \quad (5)$$

where $\mathbf{U}\equiv(u,v)$, $P$, and $\Phi$ are the velocity of the fluid, the pressure, and the electric potential, respectively; $\mathbf{E}=-\nabla\Phi$ is the electric field; $\rho$, $\eta$, and $\varepsilon$ are the mass density, dynamic viscosity, and the permittivity of the solution, respectively. $C_i$ and $\mathbf{J}_i$ are concentration and flux density of species $i$, respectively. For convenience, we use $i$=1 for K$^+$, $i$=2 for Cl$^-$, and $i$=3 for the particle. $D_i$ and $Z_i$ are the diffusion coefficient and valence of species $i$ [29], based on which the electrophoretic mobility $\mu_i$ can be calculated using the Einstein relation ($\mu_i = |Z_i D_i e/kT|$). The space charge density is given by $\rho_e = e\sum_{i=1}^{3} Z_i C_i$. The boundary conditions and numerical methods are described in ref. [25].

Figs. 2(a) and 2(b) show the time-dependent particle concentration profiles along the centerline of the channel at $V_{cm}$=15$V_T$ and $V_{cm}$=30$V_T$, respectively. A positive $V_{cm}$ causes the formation of an ion depletion zone in the vicinity of the CEMs, at the front of which the electric field is significantly amplified, extended space charges are induced, and strong vortical flows are generated (see SM S2) [13,15,16]. The amplified electrophoretic force and the vortical fluid drag force near the CEMs act jointly to prevent the particles from moving to downstream [30]. The particle trapping efficiency is therefore strongly correlated with $V_{cm}$. At a relatively low $V_{cm}$ (15$V_T$) where the electric field barrier induced by ion depletion is weak and the particles may leak into the downstream at a non-negligible rate, the particle concentration increases with a bell-shaped profile until the steady state is reached. In contrast, at $V_{cm}$=30$V_T$ where the electric field barrier is strong, the peak concentration increases until a maximum value is reached, after which the peak starts to broaden to upstream until a wide plateau of the particle concentration is formed at the steady state. Similar experimental results are shown in SM S3.

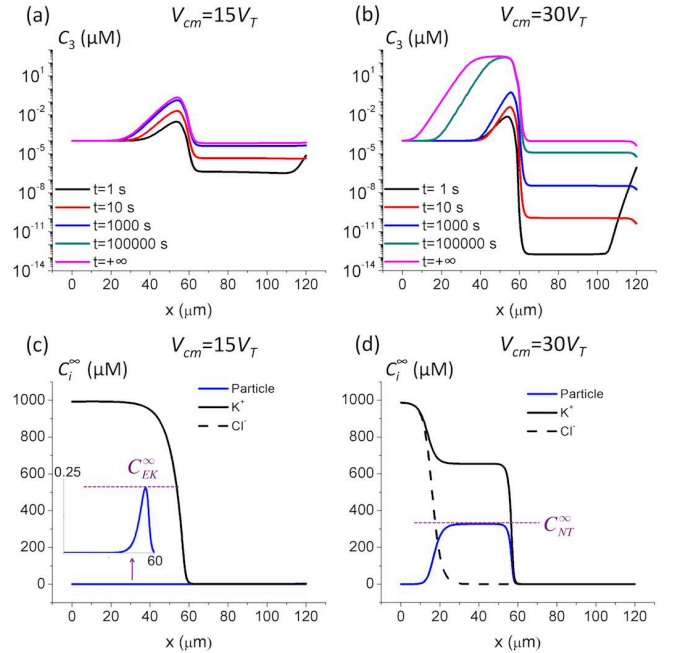

Fig. 2. (a-b) Time-dependent particle concentration profiles ($C_{3,0}$=0.1 nM) and (c-d) the steady-state concentration profiles of all charged species at $V_{cm}$=15$V_T$ and $V_{cm}$=30$V_T$, respectively.

From the steady-state concentration profiles of both the buffer ions and particles shown in Figs. 2(c) and 2(d), we can identify two distinct limits for the maximum particle concentration. At $V_{cm}$=15$V_T$, the particle concentration is significantly smaller than those of the buffer ions, so the fluid flow and electric field are dominated by the buffer ions. The maximum limiting concentration of the particle is determined by the balancing between the convective, electrophoretic, and diffusive transport of the particle. Therefore, this limit is referred to as the *electrokinetic (EK) limit* ($C_{EK}^\infty$). On the other hand, at $V_{cm}$=30$V_T$, the negatively charged particle becomes one of the majority carriers in the enrichment zone, while the corresponding concentration of Cl$^-$ is almost zero. The maximum particle concentration is limited by the value required to neutralize the positive charges carried by buffer cations. In this sense, this concentration is named the *electroneutrality (NT) limit*

($C_{NT}^\infty$).

We next derive the one-dimensional (1D) analytical expressions of the limiting EFs ($\beta^\infty$), defined as the ratio of the peak concentration to the initial concentration of the particle at the *EK* and *NT* limits ($\beta_{EK}^\infty = C_{EK}^\infty / C_{3,0}$, $\beta_{NT}^\infty = C_{NT}^\infty / C_{3,0}$). For the derivation of the *EK* limit, we only consider symmetric binary electrolytes ($D_1 = D_2 = D_{12}$, $Z_1 = -Z_2$, $C_{1,0} = C_{2,0} = C_{12,0}$), with the definitions of $a = D_{12}/D_3$, $b = Z_3/Z_2$, $b/a = \mu_3/\mu_{12}$. At the *EK* limit where $C_1 \approx C_2 \gg C_3$, the fluid velocity and electric field are defined by the buffer ions (regardless of the particle concentration). This fact permits the decoupling of the dynamics of buffer ions and that of the particle. In other words, one may solve the buffer ions distribution, fluid flow and the electric field first, based on which the distribution of the particle concentration can be further calculated [25]. Accordingly, by approximating that $J_2 \approx 0$, $J_3 \approx 0$ and applying the boundary conditions $dC_{12}^\infty/dx \approx 0$, $C_i^\infty = C_{i,0}$ at the inlet (x=0) and $C_{12}^\infty = C_{12,d}$ (downstream ion concentration after depletion) at the membrane location (x=L/2), one can obtain the distribution of the steady-state particle concentration as,

$$C_3^\infty(x) \approx C_{3,0} \cdot e^{(1-\frac{b}{a})\frac{\bar{u}}{D_3}x} \cdot \left(\frac{1-\Psi e^{\text{Pe}\frac{x}{L/2}}}{1-\Psi}\right)^b, \Psi = \left(1-\frac{C_{12,d}}{C_{12,0}}\right)e^{-\text{Pe}}. \quad (6)$$

Here, $\bar{u}$ is the average fluid velocity (net velocity), and $\text{Pe} = \bar{u}L/2D_{12}$ is the Péclet number (using L/2 as the characteristic length). Under conditions of $C_{12,d} \ll C_{12,0}$ (due to the ion depletion effect) and $e^{\text{Pe}} \gg 1$ [31], one can obtain the expression of the peak concentration ($C_{EK}^\infty$) using $dC_3^\infty/dx = 0$ at the peak and hence the EF,

$$\beta_{EK}^\infty = \frac{C_{EK}^\infty}{C_{3,0}} \approx a^{-a}b^b(a-b)^{(a-b)} \cdot e^{(a-b)\text{Pe}}. \quad (7)$$

The detailed derivation is described in SM S4.

Under the *NT* limit, the key constraint is the electroneutrality condition ($Z_1 C_1^\infty + Z_3 C_3^\infty \approx 0$, $C_2^\infty \approx 0$) at the concentration plateau. Based on this condition, by $J_1(\text{inlet}) = J_1(\text{plateau})$, $J_2 \approx 0$, $J_3 \approx 0$ and the condition $dC_i^\infty/dx \approx 0$ at both the inlet and the concentration plateau, the *NT* limit can be obtained as,

$$\beta_{NT}^\infty = \frac{C_{NT}^\infty}{C_{3,0}} \approx -\frac{Z_1}{Z_3} \cdot \frac{\mu_1/\mu_2 + 1}{\mu_1/\mu_3 + 1} \cdot \frac{C_{1,0}}{C_{3,0}}. \quad (8)$$

It should be noted that Eq. 8 is universally applicable to binary electrolytes. For symmetric binary electrolytes, Eq. 8 can be reduced to $\beta_{NT}^\infty \approx [2/(a+b)] \cdot (C_{12,0}/C_{3,0})$. The detailed derivation is described in SM S5.

Although the above analytical formulas are derived from 1D model of the electrokinetic concentration system, they enable one to extract the scaling relations between the limiting EF and critical parameters. According to Eq. 7, the *EK* limit increases exponentially with (a-b)Pe. Because Pe is proportional to $\bar{u}$, and $\bar{u}$ increases linearly with $V_{cm}$ (see SM S6 and [25]), a linear relation between $\ln \beta_{EK}^\infty$ and $V_{cm}$ must hold. This relation is demonstrated in Fig. 3(a), where approximately linear relation between the simulated $\ln \beta_{EK}^\infty$ and $V_{cm}$ is depicted. At relatively low $V_{cm}$'s, the EF is higher than that predicted by the linear relation, because the net velocity contributed by $V_L$-$V_R$ is non-negligible with weak ion depletion (see SM S6). The relation that $\ln \beta_{EK}^\infty \propto V_{cm}$ is also confirmed by experimental results in Fig. 3(b). For the *EK*-limited cases, the maximum particle concentration will eventually be capped at the *NT* limit as it increases with $V_{cm}$, so a transition from the *EK* limit to the *NT* limit is observed at sufficiently high $V_{cm}$'s. The *NT* limit is determined by the electroneutrality condition and hence independent of $V_{cm}$ (Eq. 8), as confirmed by the results in Fig. 3. Fig. 3 clearly suggests that the limiting EF follows the rule $\beta^\infty = \min\{\beta_{EK}^\infty, \beta_{NT}^\infty\}$, which is universally obeyed in the forthcoming scaling analyses.

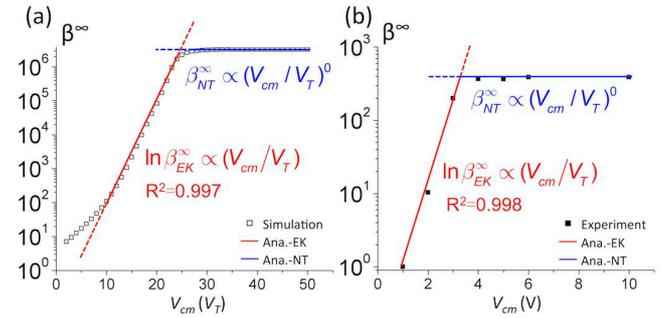

Fig. 3. The scaling relation between the steady-state enrichment factor and $V_{cm}$. (a) Comparison between the analytical solution and simulation results ($C_{3,0}$=0.1 nM). (b) Fitting between the scaling relations and experimental results ($C_{3,0}$=1 μM). Fluorescent BSA was used as the concentration target.

The electrophoretic mobility of the particle ($\mu_3$) and buffer ions ($\mu_{12}$) defines their respective transport behaviors under the electric field, which both play critical roles in determining the EF. According to Eq. 7, $\beta_{EK}^\infty$ is dominated by the term $e^{(a-b)\text{Pe}}$, based on which we have $\ln \beta_{EK}^\infty \approx (\frac{D_2}{D_3} - \frac{Z_3}{Z_2}) \cdot \frac{\bar{u}L}{2D_2}$. Considering the change of $Z_3$ (<0) with fixed $D_3=D_2/4$ and $Z_2=-1$, we have $\ln \beta_{EK}^\infty \propto (Z_3+4)$, as confirmed in Fig. 4(a). Considering the change of $D_3$ with fixed $D_2$ and $Z_3=2Z_2$, we have $\ln \beta_{EK}^\infty \propto (1/D_3 - 2/D_2)$, as confirmed in Fig. 4(b). Similarly, considering the respective changes of $Z_2$ (<0) and $D_2$, we have $\ln \beta_{EK}^\infty \propto (2/Z_2+4)$ and $\ln \beta_{EK}^\infty \propto (-2/D_2+1/D_3)$, as confirmed by the simulation data and fitting lines in Figs. 4(c) and 4(d). Generally, in the *EK*-limited regime, the EF decreases as $\mu_3(\propto |Z_3|D_3)$ increases, because the increase of leftward electrophoretic force on the particle decreases its influx into the channel. On the other hand, the EF increases as $\mu_2(\propto |Z_2|D_2)$ increases, because higher ion mobility generates stronger electric field and electroosmotic vortical flows in the ion depletion zone (see SM S7), which enhances the trapping of the particle. In the *NT*-limited regime, the EF obeys $\beta_{NT}^\infty \propto 2/(D_2/D_3+Z_3/Z_2)$ according to Eq. 8, which weakly depends on $\mu_3$ and $\mu_2$, as confirmed in Figs. 4(a-d). Finally, as shown in Figs. 4(a) and 4(b), Eq. 7 does not apply to the cases with too small $\mu_3$, where the trapping of the particle is compromised by the weak electrical force, and the condition of $J_3 \approx 0$ is no longer valid.

The analytical solutions in Eqs. 7 and 8 can also be used to elucidate the effects of initial particle concentration. At the *EK* limit where $C_1 \approx C_2 \gg C_3$, the properties of the system do not change with the particle concentration, so the limiting EF in Eq. 7 is independent of the initial particle concentration, i.e. $\beta_{EK}^\infty \propto (C_{3,0})^0$, as confirmed by the simulation and experimental results in Figs. 5 and S8-1(a). On the other hand, at the *NT* limit, the particle concentration ($C_{NT}^\infty$) is a constant solely determined by the electroneutrality condition. Therefore, the limiting EF is reversely proportional to the initial concentration, i.e. $\beta_{NT}^\infty \propto (C_{3,0})^{-1}$, as confirmed by Eq. 8 and Figs. 5 and S8-1(b).

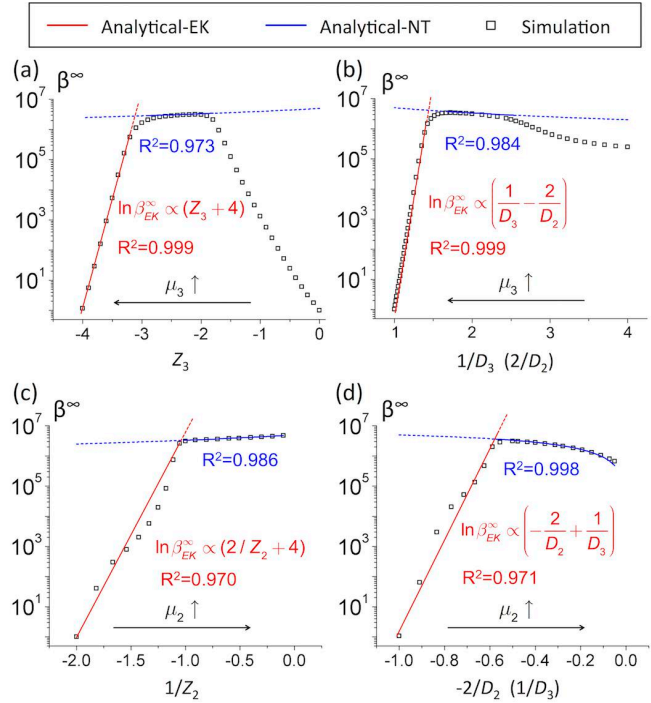

Fig. 4. The scaling relations between the steady-state enrichment factor and the electrophoretic mobility of the particle and buffer ions. (a) Effect of the valence of the particle. (b) Effect of the diffusion coefficient of the particle. (c) Effect of the valence of cation. (d) Effect of the diffusion coefficient of buffer ions. ($V_{cm}=30V_T$, $C_{3,0}=0.1$ nM).

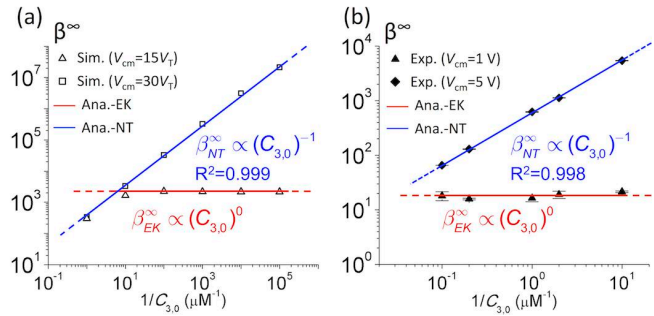

Fig. 5. The scaling relations between the steady-state EF and initial particle concentration. (a) Fitting between the scaling relations and simulation results. (b) Fitting between the scaling relations and experimental results. Fluorescent ssDNA was used as the concentration target.

The buffer concentration ($C_{12,0}$) strongly affects the formation of the ion depletion zone, which has profound effects on the EF. At low buffer ion concentrations where ion depletion is sufficiently developed, the strong electrical barrier is maintained, which renders the system in the *NT*-limited regime. According to Eq. 8, the limiting EF is proportional to $C_{12,0}$, i.e. $\beta_{NT}^\infty \propto C_{12,0}$, as confirmed by both the simulation and experimental results in Fig. 6. In contrast, due to the limited transport capacity of the CEMs, ion depletion is weakly developed at high buffer concentrations. Therefore, the system lies in the *EK*-limited regime, with low EFs being achieved. In this regime, the fluid flow is dominated by the quasi-equilibrium EOF induced by the tangential electric field and charges in the electric double layers, which is proportional to $(C_{12,0})^{-0.5}$ [32]. According to Eq. 7, the limiting EF is proportional to $(C_{12,0})^{-0.5}$, i.e. $\beta_{EK}^\infty \propto (C_{12,0})^{-0.5}$, as confirmed by both the simulation and experimental results in Fig. 6. This explains the difficulty in the concentration of particles in high concentration buffers.

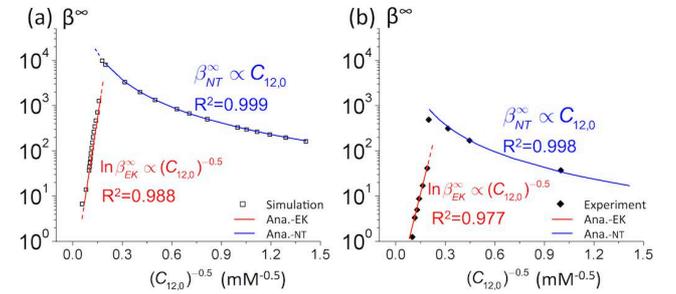

Fig. 6. The scaling relations between the steady-state EF and buffer concentration. (a) Fitting between the scaling relations and simulation results ($C_{3,0}=1$ μM). (b) Fitting between the scaling relations and experimental results ($C_{3,0}=1$ μM). Fluorescent ssDNA was used as the concentration target.

In this letter, we report the first theoretical model of ICP-based electrokinetic concentration. Relations between the enrichment factor and critical parameters are analytically elaborated for the two dissimilar operating regimes (*EK* and *NT* limits), providing explanations on the qualitatively different experimental behaviors observed in the past. These results significantly advance the scientific understanding of electrokinetic concentration with implications in many electrochemical systems.

This work is supported by the National Natural Science Foundation of China (Grant Nos. 11372229, 21576130, 21490584) and the National Institutes of Health of the United States (Grant No. U19AI109755).

*Correspondence: lizirui@gmail.com; jyhan@mit.edu.*

# Supplementary Material

# Theoretical Limits and Scaling Laws for Electrokinetic Molecular Concentration via Ion Concentration Polarization

Wei Ouyang,[1] Zirui Li,[2,*] Xinghui Ye,[2] Jongyoon Han[1,2,3,*]

[1]*Department of Electrical Engineering and Computer Science, Massachusetts Institute of Technology, Cambridge, Massachusetts, 02139, USA.*

[2]*Institute of Laser and Optoelectronic Intelligent Manufacturing, College of Mechanical and Electrical Engineering, Wenzhou University, Wenzhou, 325035, P.R. China.*

[3]*Department of Biological Engineering, Massachusetts Institute of Technology, Cambridge, Massachusetts, 02139, USA.*

## S1. Experimental details

The microchannels were fabricated by standard PDMS soft lithography as described in ref. [1]. The two parallel microchannels are 2 cm long, 100 μm wide and 15 μm deep with a gap of 250 μm. A thin strip of Nafion membrane was patterned on a glass substrate using the microflow patterning technique [2]. 5 wt.% Nafion® perfluorinated resin solution in lower aliphatic alcohols and water (Sigma-Aldrich, 274704-25ML) was used in this work. The PDMS microchannel for depositing the Nafion membrane was 200 μm wide and 20 μm deep. The final Nafion membrane thickness after curing was ~1 μm. Finally, the PDMS microchannel was bonded to the Nafion-patterned glass substrate via plasma bonding using the factory settings (Femto Science Covance).

In the concentration experiments, we used 10 mM KCl solution as the buffer. The analyte molecule is Alexa Fluor 488-labeled bovine serum albumin (BSA) and Alexa Fluor 488-labeled 21-base ssDNA (5'-AGTCAGTCAGTCAGTCAGTCA-3'). An inverted fluorescent microscope (IX71, Olympus) and a CCD camera (Sensicam qe, Cook Corp.) were used for imaging. A mechanical shutter was used to reduce the photo-bleaching effect. Micro-manager (www.micromanager.org) was used to synchronize the CCD camera and the mechanical shutter. ImageJ (National Institutes of Health, USA) was used for image analysis. A DC power supply (Stanford Research Systems, Sunnyvale, CA) was used to apply the voltages. A multi-meter (6514, Keithley Inc.) was used to measure the voltages. Ag/AgCl electrodes (A-M Systems Inc.) were used as electrodes.

## S2. Fluid velocity, electric field, and flux of the particle near the CEM

An electrical configuration with $V_{cm}>0$ creates the normal electric field ($E_N$) that drives the selective transport of cations through the CEMs. The depletion of cations results in significantly amplified electric field near the CEMs (Poisson equation), which repulses anions from this region, forming the ion depletion zone. At sufficiently high $E_N$, extended space charges are induced near the CEMs, which facilitates the generation of a convective EOF near the CEMs that is much faster than the primary electroosmosis (of the first kind). As shown in Fig. S2-1(a), the fluid velocity close to the CEM is as fast as 12.5 mm/s. Vortical flows are generated in the vicinity of the CEMs such that the incompressibility of fluid (continuity of fluid flow) can be maintained. As the electric field distribution in Fig. S2-1(b) implies, the strong tangential electric force at the front of the ion depletion zone, and the normal electric force near the CEMs that sends the particles towards the backflow of the fluid act jointly to prevent the particles from moving to downstream, which is the basis of the electrokinetic concentration effect. Fig. S2-1(c) shows the flux of the particle near the vortical flow. Instead of following the fluid streamlines to downstream, the particles are recirculated to upstream by the backflow of the fluid. The mismatch between the particle flux entering the channel and leaving the channel leads to the continuous concentration of the particle until the steady state is reached, as indicated by the particle concentration map in Fig. S2-1(d).

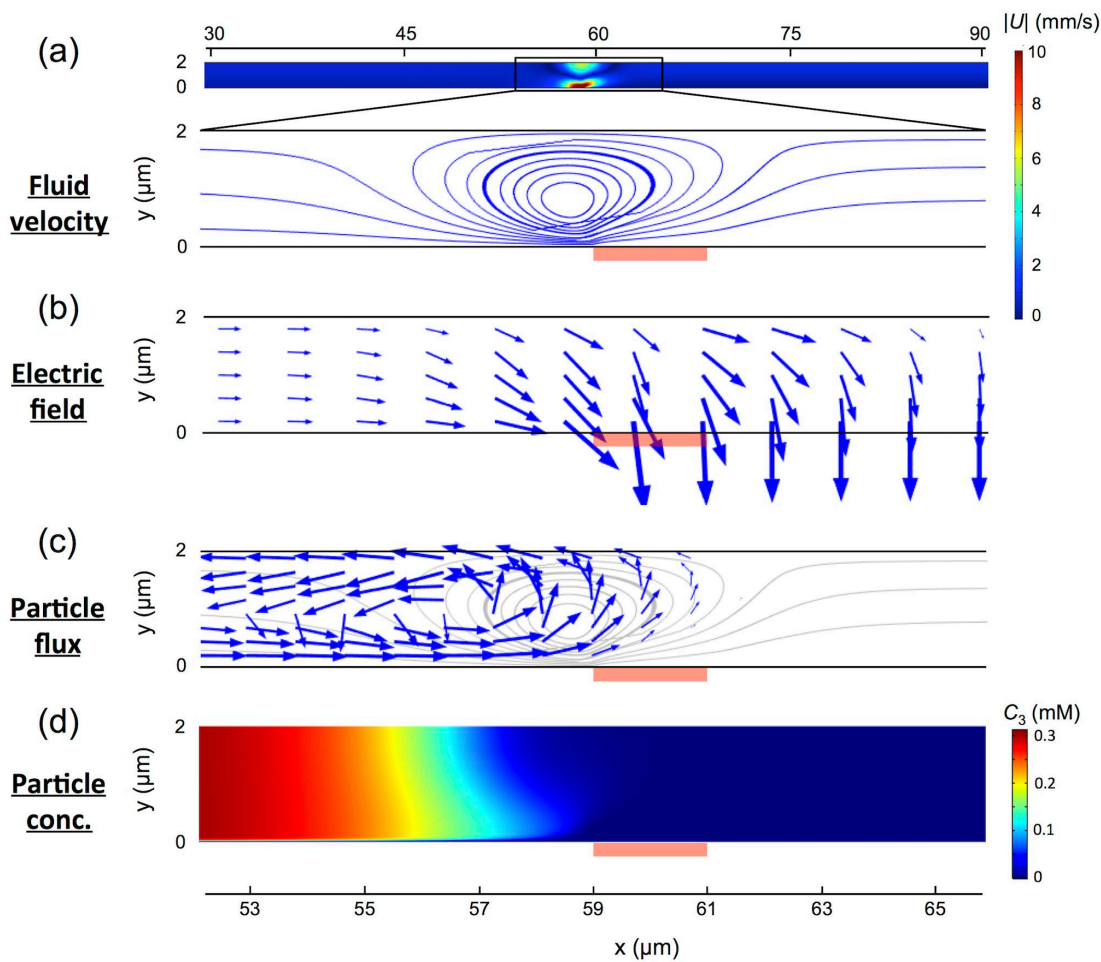

**Fig. S2-1.** (a) Fluid velocity, (b) electric field, (c) the particle flux density, (d) the particle concentration near the CEM at the steady state at $V_{cm}=30V_T$. The lengths of arrows are plotted in natural logarithm-scale. The light grey fluid streamlines are plotted as a reference. Only the lower half of the channel is plotted.

## S3. Experimental observation of the *EK* and *NT* limits

In this part, we show the typical experimental concentration behaviors in the *EK*-limited and *NT*-limited regimes. For the data shown, we used $V_1=20$ V, $V_2=10$ V, $V_3=13.5$ V to render the concentrator in the *EK*-limited regime, and $V_1=20$ V, $V_2=10$ V, $V_3=10$ V to render the concentrator in the *NT*-limited regime. For the *EK* limit, the initial DNA concentration is 100 nM, achieving a near-steady-state enrichment factor of ~300 in 9000 s. Because it generally takes longer time to reach the *NT* limit compared to the *EK* limit, we used a higher initial DNA concentration (1 μM) to shorten the time to the *NT* limit, with a near-steady-state enrichment factor of ~600 achieved in 6000 s. The evolution of the fluorescence profiles are shown in Fig. S3-1.

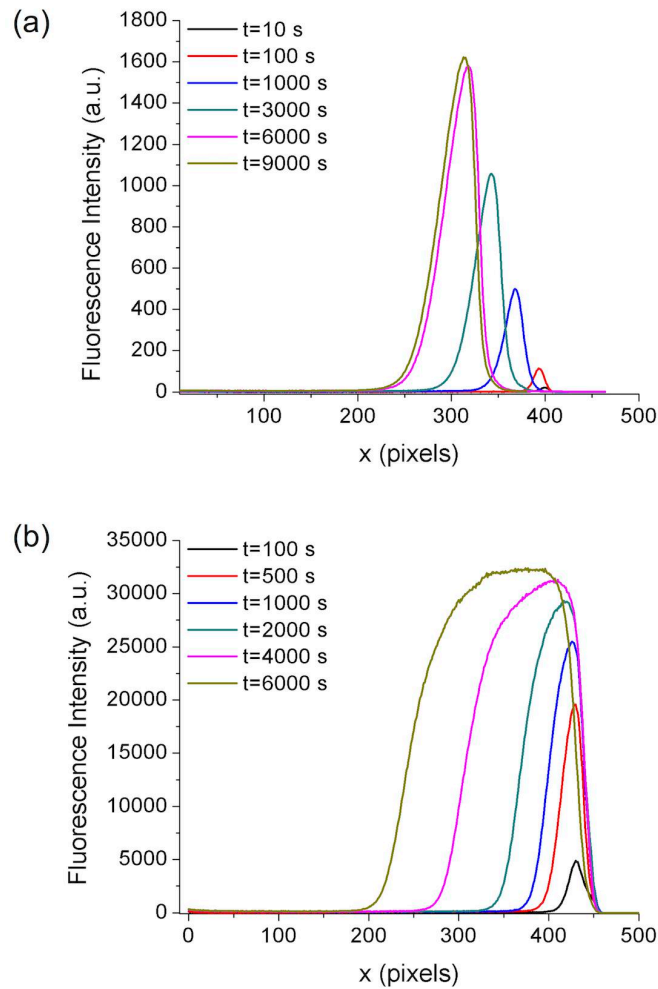

**Fig. S3-1.** Evolution of the fluorescence profiles (a) in the *EK*-limited regime and (b) in the *NT*-limited regime.

## S4. Derivation of the *EK* limit

The basic idea is that, at the *EK* limit, the concentration of the particle is significantly smaller than those of the buffer ions ($C_1 \approx C_2 \gg C_3$), so the electric field and fluid velocity are defined by the buffer ions (regardless of the particle concentration). One can solve the distribution of the buffer ions, the electric field, and fluid velocity first, based on which the distribution of the particle concentration can be further calculated. In this derivation, we will only consider symmetric binary electrolytes (e.g. KCl) as the buffer, i.e. $D_1 = D_2 = D_{12}$, $Z_1 = -Z_2$, $C_{1,0} = C_{2,0} = C_{12,0}$.

Because the electrophoretic force is co-directional with the fluid drag force for the cation and counter-directional for the anion and particle, and the concentration of the anion is much greater than that of the particle, the fluxes of charged species satisfy $J_1 \gg J_2 \gg J_3$, as supported by the simulation data in Fig. S4-1.

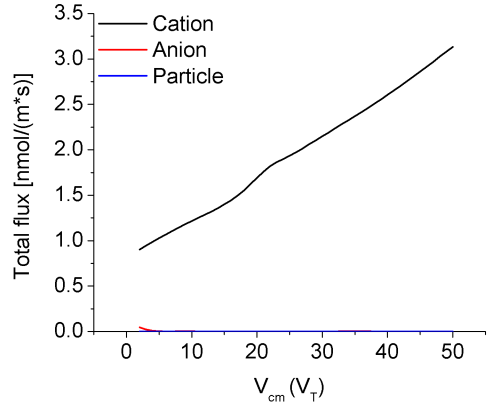

**Fig. S4-1.** The fluxes of the cation, anion and particle at $x=30$ μm at the steady state at different $V_{cm}$'s. Note that at the steady state, the fluxes are constant along the *x*-axis.

By approximating $J_2 \approx 0$ and $J_3 \approx 0$, the fluxes of the cation, anion, and particle can be written as,

$$J_1 = -D_{12}\frac{dC_{12}}{dx} + (\bar{u} + \mu_{12}E)C_{12} = J_C \quad \text{S4.1}$$

$$J_2 = -D_{12}\frac{dC_{12}}{dx} + (\bar{u} - \mu_{12}E)C_{12} \approx 0 \quad \text{S4.2}$$

$$J_3 = -D_3\frac{dC_3}{dx} + (\bar{u} - \mu_3 E)C_3 \approx 0 \quad \text{S4.3}$$

Note that $J_C$ is a constant along the pre-CEM channel at the steady state, and $\bar{u}$ is a constant along the channel due to the incompressibility of fluid.

Subtracting Eq. S4.2 from Eq. S4.1 yields that,

$$\mu_{12} E C_{12} \approx J_C / 2 . \tag{S4.4}$$

Substituting Eq. S4.4 to Eq. S4.2 yields that,

$$-D_{12} \frac{dC_{12}}{dx} + \bar{u} C_{12} - \frac{J_C}{2} \approx 0 . \tag{S4.5}$$

At the inlet ($x=0$), $\frac{dC_{12}}{dx}\big|_{x=0} \approx 0$ and $C_{12}(x=0) = C_{12,0}$, which leads to,

$$J_C \approx 2 \bar{u} C_{12,0} . \tag{S4.6}$$

Substituting Eq. S4.6 to Eq. S4.5 yields that,

$$-D_{12} \frac{dC_{12}}{dx} + (C_{12} - C_{12,0}) \bar{u} \approx 0 . \tag{S4.7}$$

Therefore, $C_{12}(x) \approx C_{12,0} + A e^{\frac{\bar{u}}{D_{12}} x}$, where $A$ is a constant.

We define the downstream ion concentration (after depletion) as $C_{12,d}$. At the membrane ($x = L/2$), we have $C_{12}(x = L/2) = C_{12,d}$, which leads to,

$$C_{12}(x) \approx C_{12,0} \left[ 1 + \left( \frac{C_{12,d}}{C_{12,0}} - 1 \right) e^{\frac{\bar{u}}{D_{12}}(x - \frac{L}{2})} \right] . \tag{S4.8}$$

Substituting Eq. S4.6 and Eq. S4.8 to Eq. S4.4 yields that,

$$E(x) \approx \frac{\bar{u}}{\mu_{12}} \cdot \frac{1}{1 + \left( \frac{C_{12,d}}{C_{12,0}} - 1 \right) e^{\frac{\bar{u}}{D_{12}}(x - \frac{L}{2})}} . \tag{S4.9}$$

Up to this point, the distributions of buffer ions and the electric field have been solved.

Next, by substituting Eq. S4.9 to Eq. S4.3 and applying the boundary condition $C_3(x=0) = C_{3,0}$, one can obtain the distribution of the particle concentration along the x-axis as,

$$C_3(x) \approx C_{3,0} e^{\left( \frac{\bar{u}(\mu_{12} - \mu_3)}{\mu_{12} D_3} \right) x} \left( \frac{1 - \Psi e^{\frac{\bar{u}}{D_{12}} x}}{1 - \Psi} \right)^{\frac{D_{12} \mu_3}{D_3 \mu_{12}}} , \quad \Psi = \left( 1 - \frac{C_{12,d}}{C_{12,0}} \right) e^{-\frac{\bar{u}}{D_{12}} \frac{L}{2}} . \tag{S4.10}$$

At the concentration peak ($C_{EK}^{\infty}$), $\frac{dC_3(x)}{dx} = 0$, from which one can obtain that,

$$C_{EK}^{\infty} \approx C_{3,0} \left[ \frac{1}{\Psi}(1-\frac{\mu_3}{\mu_{12}}) \right]^{\frac{D_{12}}{D_3}(1-\frac{\mu_3}{\mu_{12}})} \left( \frac{1}{1-\Psi} \frac{\mu_3}{\mu_{12}} \right)^{\frac{D_{12}\mu_3}{D_3 \mu_{12}}} . \qquad \text{S4.11}$$

Under conditions of $C_{12,0} \gg C_{12,\text{dep}}$ (due to the ion depletion effect) and $\bar{u}L/2D_{12} \gg 1$, Eq. S4.11 can be simplified to,

$$C_{EK}^{\infty} \approx C_{3,0} \left( 1 - \frac{\mu_3}{\mu_{12}} \right)^{\frac{D_{12}}{D_3}(1-\frac{\mu_3}{\mu_{12}})} \left( \frac{\mu_3}{\mu_{12}} \right)^{\frac{D_{12}\mu_3}{D_3 \mu_{12}}} \left( e^{\frac{\bar{u}}{D_{12}} \frac{L}{2}} \right)^{\frac{D_{12}}{D_3}(1-\frac{\mu_3}{\mu_{12}})} . \qquad \text{S4.12}$$

Defining $a = \frac{D_{12}}{D_3}$, $b = \frac{Z_3}{Z_2}$, $\frac{b}{a} = \frac{\mu_3}{\mu_{12}}$, and the Péclet number $\text{Pe} = \frac{\bar{u}}{D_{12}} \cdot \frac{L}{2}$, the EF under the

*EK* limit can be re-written as,

$$\beta_{EK}^{\infty} = \frac{C_{EK}^{\infty}}{C_{3,0}} \approx a^{-a} b^b (a-b)^{(a-b)} \cdot e^{(a-b)\cdot\text{Pe}} . \qquad \text{S4.13}$$

## S5. Derivation of the *NT* limit

At the steady state, the fluxes of all charged species are constant along the *x*-axis. Considering the inlet and the concentration plateau, one can have $J_1(0) = J_1(p)$, $J_2(0) = J_2(p) \approx 0$, $J_3(0) = J_3(p) \approx 0$, where "0" denotes the inlet, and "p" denotes the concentration plateau. These equations can be expanded to,

$$-- D_1 \frac{dC_1(0)}{dx} + (\bar{u} + \mu_1 E_p) C_1(0) = -D_1 \frac{dC_1(p)}{dx} + (\bar{u} + \mu_1 E_p) C_1(p), \quad \text{S5.1}$$

$$-D_2 \frac{dC_2(0)}{dx} + (\bar{u} - \mu_2 E_p) C_2(0) = -D_2 \frac{dC_2(p)}{dx} + (\bar{u} - \mu_2 E_p) C_2(p) \approx 0, \quad \text{S5.2}$$

$$-D_3 \frac{dC_3(0)}{dx} + (\bar{u} - \mu_3 E_p) C_3(0) = -D_3 \frac{dC_3(p)}{dx} + (\bar{u} - \mu_3 E_p) C_3(p) \approx 0. \quad \text{S5.3}$$

At the inlet, the concentrations of all charged species equal the initial concentration, and the concentration gradients can be considered zero. At the concentration plateau, the concentration gradients of all charged species are zero. Therefore, $C_i(0) = C_{i,0}$, $\frac{dC_i(0)}{dx} = 0$, $\frac{dC_i(p)}{dx} = 0$, *i*=1, 2, 3. Eqs. S5.1-S5.3 can be reduced to,

$$(\bar{u} + \mu_1 E_0) C_{1,0} = (\bar{u} + \mu_1 E_p) C_1(p), \quad \text{S5.4}$$

$$(\bar{u} - \mu_2 E_0) C_{2,0} = (\bar{u} - \mu_2 E_p) C_2(p) \approx 0, \quad \text{S5.5}$$

$$(\bar{u} - \mu_3 E_0) C_{3,0} = (\bar{u} - \mu_3 E_p) C_3(p) \approx 0. \quad \text{S5.6}$$

In Eq. S5.5, the total flux is zero, and the anion concentration at the inlet is much greater than zero, one can have,

$$\bar{u} \approx \mu_2 E_0. \quad \text{S5.7}$$

In Eq. S5.6, the total flux is zero, and the particle concentration at the plateau is much greater than zero, one can have,

$$\bar{u} \approx \mu_3 E_p. \quad \text{S5.8}$$

From Eq. S5.7 and Eq. S5.8, one can have,

$$E_p \approx \frac{\mu_2}{\mu_3} E_0. \quad \text{S5.9}$$

Substituting Eq. S5.7 and Eq. S5.9 to Eq. S5.4 yields that,

$$C_1(p) \approx \frac{\mu_1/\mu_2 + 1}{\mu_1/\mu_2 + 1} C_{1,0}. \quad \text{S5.10}$$

At the concentration plateau, $C_2(p) \approx 0$, so the electroneutrality condition requires that,

$$Z_1 C_1(p) + Z_3 C_3(p) \approx 0. \quad \text{S5.11}$$

Combining Eq. S5.10 and Eq. S5.11 gives that,

$$C_{NT}^{\infty} = C_3(\text{p}) \approx -\frac{Z_1}{Z_3} \cdot \frac{\mu_1/\mu_2 + 1}{\mu_1/\mu_3 + 1} \cdot C_{1,0}.$$ S5.12

Therefore, the EF at the *NT* limit is,

$$\beta_{NT}^{\infty} = \frac{C_{NT}^{\infty}}{C_{3,0}} \approx -\frac{Z_1}{Z_3} \cdot \frac{\mu_1/\mu_2 + 1}{\mu_1/\mu_3 + 1} \cdot \frac{C_{1,0}}{C_{3,0}}.$$ S5.13

Note that this equation is generally applicable to arbitrary binary electrolytes. For symmetric electrolytes ($D_1 = D_2 = D_{12}, Z_1 = -Z_2, C_{1,0} = C_{2,0} = C_{12,0}$), Eq. S5.14 can be reduced to,

$$\beta_{NT}^{\infty} \approx \frac{2}{a+b} \cdot \frac{C_{12,0}}{C_{3,0}}.$$ S5.14

## S6. The relation between $\bar{u}$ and $V_{cm}$

The velocity profile of the system is shown in Fig. S6-1. The action of a tangential electric field upon the induced space charges near the CEMs induces a non-equilibrium electroosmotic flow, which is named the electroosmosis of the second kind (EOF2) by Dukhin *et. al* [3]. This electroosmotic slip is much faster (>10x) than the primary electroosmosis (EOF1) in the bulk channel. Consequently, a pair of vortices is generated near the CEMs to satisfy the incompressibility of fluid. At the same time, a pressure-driven flow is induced in the bulk channel that speeds up the net fluid velocity ($\bar{u}$), as indicated by the parabolic flow profiles at higher $V_{cm}$'s.

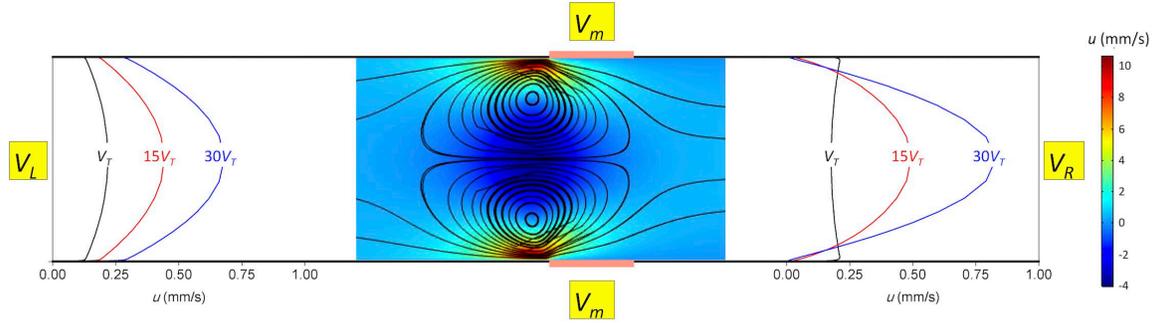

**Fig. S6-1** Fluid velocity profiles in the channel. The electrical configuration is $V_{LR}=20V_T$, with $V_{cm}$ labeled on each curve. The color map represent the x-direction velocity (*u*) at $V_{cm}=20V_T$.

Theoretical modeling of the fluid flow of the system is challenging, as it involves the coupling of EOF1 and EOF2. Rubinstein *et. al* [4-7] and Kim *et. al* [8] studied the symmetric case of the system ($V_L=V_R$), in which there is no net tangential fluid flow ($\bar{u}=0$). According to their studies, the EOF2 slip velocity ($u_{s2}$) near the CEM surface is proportional to the square or cube of $V_{cm}$, depending on the magnitude of the electric field applied. This nonlinear dependence is observed in our model by setting $V_L=V_R$, as shown in Fig. S6-2.

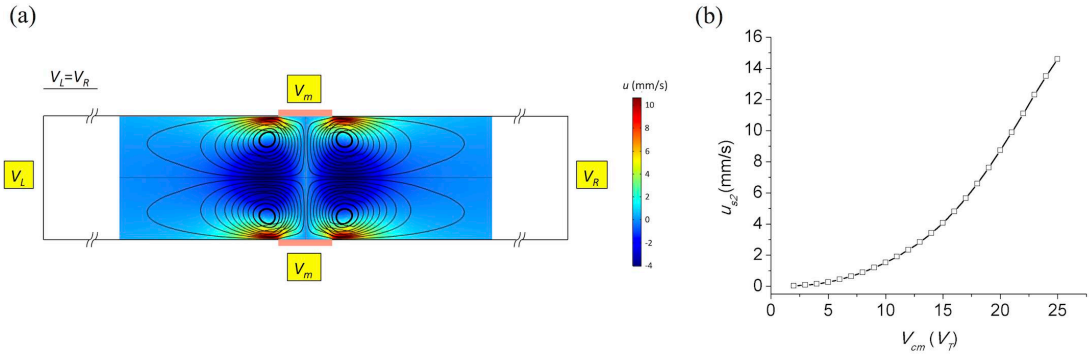

**Fig. S6-2.** (a) Velocity streamlines and velocity magnitude in a symmetric system. The color map represent the x-direction velocity ($u$) at $V_{cm}=21V_T$. (b) The EOF2 slip velocity ($u_{s2}$) in the vicinity of the CEMs (x=59 μm).

However, when the symmetry is broken ($V_{LR}=V_L-V_R>0$), EOF2 contributes to the net tangential fluid flow. Fig. S6-3(a) shows the velocity profiles along the cross-section of the channel in the vicinity of the CEMs (x=59 μm) at different $V_{cm}$'s. We take the maximum velocity as the EOF2 slip velocity ($u_{s2}$). As shown in Fig. S6-3(b), $u_{s2}$ is proportional to $V_{cm}$, which is clear departure from the scaling relation in the symmetric scenario. The detailed mechanism calls for further theoretical studies.

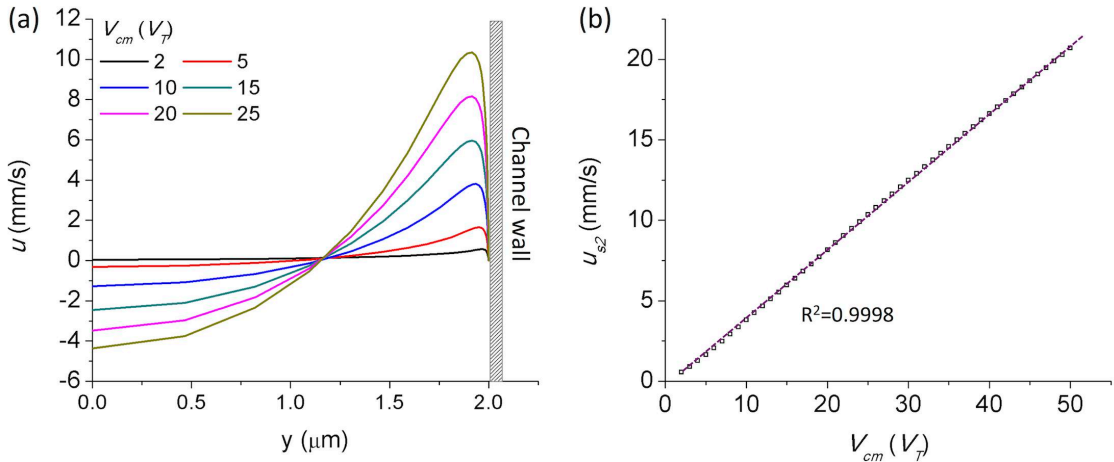

**Fig. S6-3.** (a) The x-direction velocity ($u$) profiles along the cross-section of the channel in the vicinity of the CEMs (x=59 μm) at different $V_{cm}$'s at $V_{LR}=15V_T$. (b) Dependence of the EOF2 slip velocity ($u_{s2}$) on $V_{cm}$ at $V_{LR}=15V_T$.

Fig. S6-4 shows the dependence of $\bar{u}$ on $V_{cm}$ and $V_{LR}$. When $V_{cm}$ is relatively low compared to $V_{LR}$, the system is dominated by the tangential electric field set up by $V_{LR}$. In this regime, $\bar{u}$ is mainly determined by EOF1, because EOF2 is relatively small due to the weak ion depletion effect. Consequently, $\bar{u}$ increases approximately linearly with $V_{cm}$ at a small slope, as lowering $V_m$ also increases the upstream tangential electric field that drives EOF1. As $V_{cm}$ further increases, the system becomes dominated by the electric field set up by $V_L$ and $V_m$. In this regime, $\bar{u}$ is mainly determined by EOF2. Because EOF2 is linear with $V_{cm}$ as aforementioned and EOF2 is much faster than EOF1, $\bar{u}$ increases linearly with $V_{cm}$ with a large slope. Lastly, the higher $V_{LR}$ is, the higher $V_{cm}$ is needed to enter the EOF2-dominated regime.

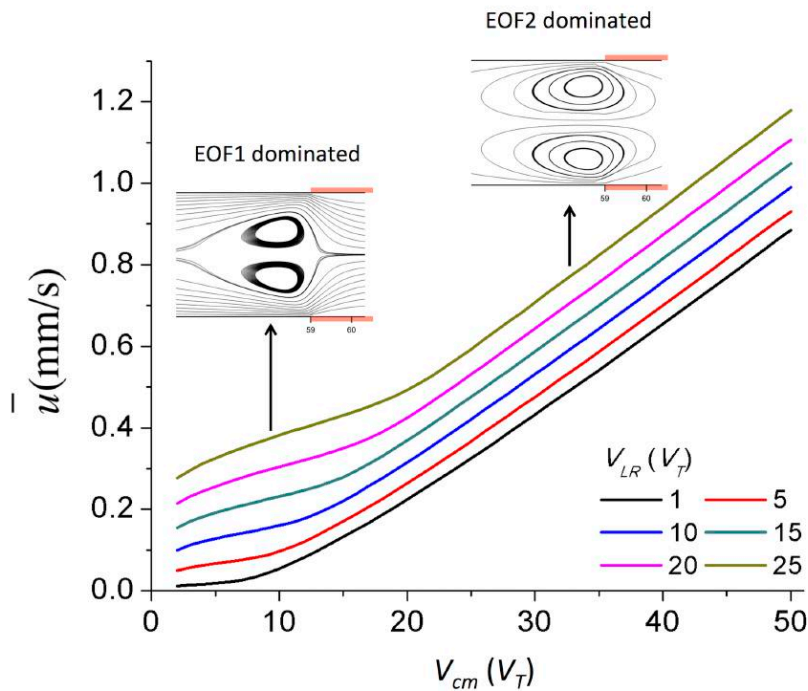

**Fig. S6-4.** Dependence of the net fluid velocity $\bar{u}$ on $V_{LR}$ and $V_{cm}$. When $V_{cm}$ is small, $\bar{u}$ is dominated by EOF1, which is approximately linear with $V_{cm}$. When $V_{cm}$ is large, $\bar{u}$ is dominated by EOF2, which is linear with $V_{cm}$ with a higher slope than that of the EOF1-dominated regime.

## S7. Effect of the electrophoretic mobility of buffer ions

High electrophoretic mobility of buffer ions ($\mu_{12}$) accelerates the transport of cations through the ion depletion zone and the CEM, and the repulsion of anions from the ion depletion zone, thereby forming ion depletion zones with lower ion concentrations, as shown in Fig. S7-1(a). Consequently, thicker extended space charge layers are formed in the ion depletion zone at higher $\mu_{12}$, as indicated by the color maps in Fig. S7-1(b). According to the Poisson equation, more abrupt changes of the electric field exist in the ion depletion zone at higher $\mu_{12}$, which leads to stronger electric fields, as indicated by the arrows in Fig. S7-1(b). As a result, the non-equilibrium EOF in the ion depletion zone is accelerated and the trapping of the particles is enhanced, leading to higher EFs.

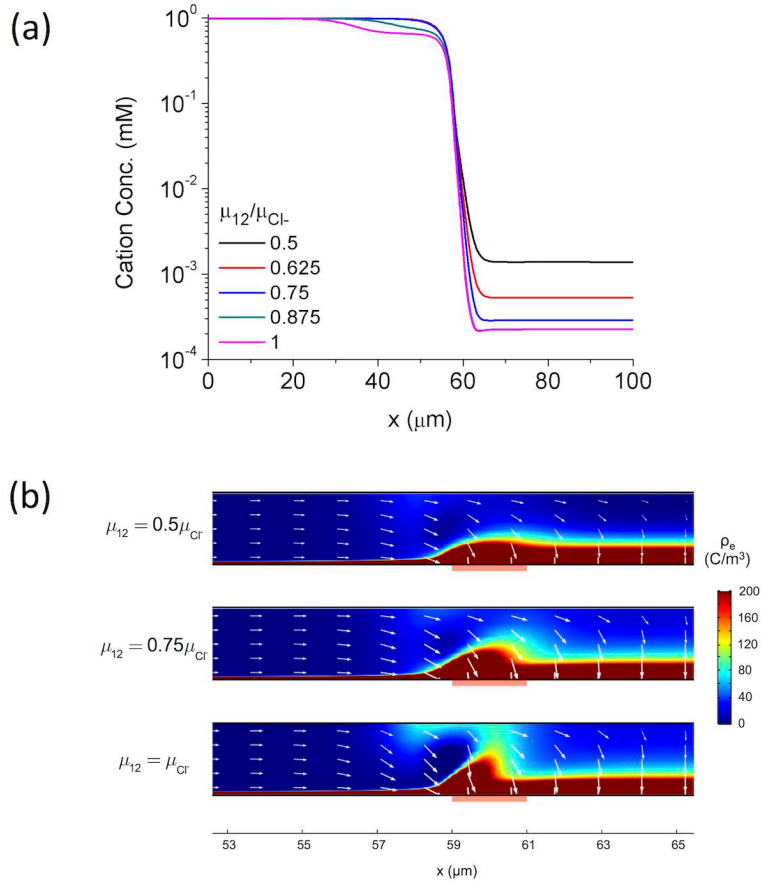

**Fig. S7-1.** Effect of the electrophoretic mobility of buffer ions. (a) Cation concentration profiles at different $\mu_{12}$. (b) The space charge density (color map) and the electric field (arrows plotted in natural logarithm-scale at different $\mu_{12}$.

## S8. Time-dependent profiles of the particle concentration at different initial particle concentrations

To study how the initial particle concentration ($C_{3,0}$) affects $\beta^\infty$, we analyzed the concentration behaviors at different initial concentrations (Fig. S8-1). At $V_{cm}=15V_T$, for initial concentrations of $10^{-5}$ µM to $10^{-1}$ µM, the steady-state particle concentrations scale proportionally with their respective initial concentrations during the whole course (Fig. S8-1(a)), because the electric field and fluid flow remain the same when the concentrated particle is still a minority carrier. The maximum concentration corresponds to the *EK* limit. Beyond $C_{3,0}=1$ µM, the maximum concentration is capped near the *NT* limit, before reaching the *EK* limit. In contrast, at $V_{cm}=30V_T$, all the curves approach a single limit at a significantly different times regardless of the initial concentrations (Fig. S8-1(b)), which corresponds to the *NT* limit. In short, the steady-state peak concentration of the particle follows the rule: $C_{3,\max}^\infty = \min\{C_{EK}^\infty, C_{NT}^\infty\}$.

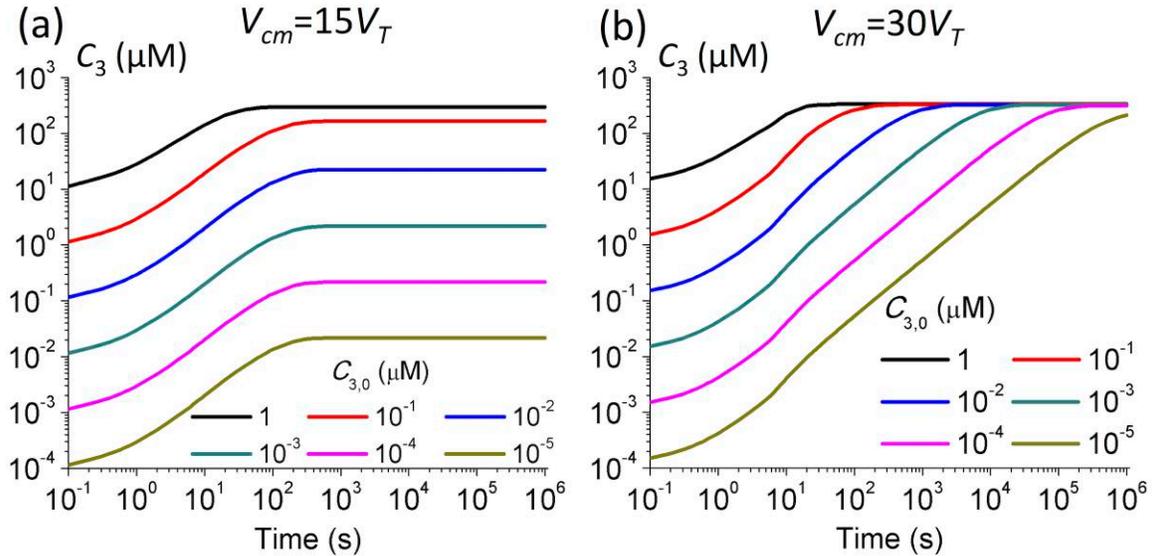

**Fig. S8-1.** Time-dependent profiles of the particle concentrations at different initial concentrations by simulation (a) at $V_{cm}=15V_T$ and (b) at $V_{cm}=30V_T$ ($V_{LR}=20V_T$).